\documentclass[manuscript]{aastex}

\shorttitle{A Mechanism of Exciting}
\shortauthors{Chen et al.}

\begin{document}


\title{A Mechanism of Exciting Planetary Inclination and Eccentricity \\
through a Residual Gas Disk}


\author{Yuan-Yuan Chen, Hui-Gen Liu, Gang Zhao, Ji-Lin Zhou }



\email{cyy198531@nju.edu.cn, zhoujl@nju.edu.cn}

\altaffiltext{}{School of Astronomy and Space Science \& Key Laboratory of Modern
Astronomy and Astrophysics in Ministry of Education, Nanjing
University, Nanjing, China, 210093}

\begin{abstract}

Accordling to the theory of Kozai resonance, the initial mutual
inclination between a small body and a massive planet in an outer
circular orbit is as high as $\sim39.2^{\circ}$ for pumping the
eccentricity of the inner small body. Here we show that, with the
presence of a residual gas disk outside two planetary orbits, the
inclination can be reduced as low as a few degrees. The presence of
disk changes the nodal precession rates and directions of the planet
orbits. At the place where the two planets achieve the same nodal
processing rate, vertical secular resonance would occur so that
mutual inclination of the two planets will be excited, which might
trigger the Kozai resonance between the two planets further.
However, in order to pump an inner Jupiter-like planet, the
conditions required for the disk and the outer planet are relatively
strict. We develop a set of evolution equations, which can fit the
N-body simulation quite well but be integrated within a much shorter
time. By scanning the parameter spaces using the evolution
equations, we find that, a massive planet ($10M_J$) at $30$AU with
$6^o$ inclined to a massive disk ($50M_J$) can finally enter the
Kozai resonance with an inner Jupiter around the snowline. And a
$20^{\circ}$ inclination of the outer planet is required for
flipping the inner one to a retrograde orbit. In multiple planet
systems, the mechanism can happen between two nonadjacent planets,
or inspire a chain reaction among more than two planets. This
mechanism could be the source of the observed giant planets in
moderate eccentric and inclined orbits, or hot-Jupiters in close-in,
retrograde orbits after tidal damping.

\end{abstract}





\section{Introduction}

Kozai mechanism is a kind of secular effect that occurred in
hierarchical three-body systems\citep{Lid62,Koz62,Naoz11}. In the
limit of circular restricted three-body model, the test particle in
an inner orbit can be pumped to a highly eccentric or inclined orbit
as long as its initial inclinations relative to the outer massive
perturber is $\ge 39.2^o$\citep{Koz62,Inn97}. Further more,
\citet{LN11} show that, when the massive perturber is in an
eccentric orbit, the effect of octupole terms in the perturbing
function will be effective so that the inner test particle may be
repelled to retrograde orbits relative to the massive planet orbit.

One of the most prominent applications for Kozai mechanism is on the
formation of orbital configurations for hot Jupiters (HJ). Recent
observations of Rossiter-McLaughlin (RM) effect \citep{Rossiter24,
McLaughlin24} show that most of the HJs might be in orbits
misaligned with stellar spins. Actually, for the 53 HJs with RM
effect measurements, at least 8 HJs might be in retrograde
motions\citep{Winn10, Triaud10, Brown12, Alb12}. As the classical
core accretion scenario says that planets were formed in a
protoplanetary disk surrounding the protostar, the existence of HJs
in highly inclined orbits infers that some dynamical mechanisms to
pump their inclinations must be exist after their formation. As the
so-called disk migration scenario \citep{Lin86, LBP96} failed to
explain the existence of HJs in retrograde orbits, Kozai mechnism
was invoked to excite the orbital inclinations\citep{Wu03,FT07,
Nao11a, Nag08}.

\citet{Wu03,FT07} proposed that a third massive body (either a
binary or a brown dwarf companion) with a high orbital
inclination($\ge39.2^{\circ}$) can trigger the Kozai resonance so
that the orbital eccentricity of inner planets can be pumped up to
near 1, which can be damped at periastron of the orbit, with its
excited inclination being preserved. However, the population studies
establish that only $10\%$ of HJs can be explained by Kozai
migration due to binary companions \citep{Wu07}, while most of the
HJ systems did not find any stellar or substellar companions. An
alternative choice is whether the outer perturber can be replaced by
a massive planet. Although this is possible, a very high mutual
inclination between the two planets is required. E.g.,
\citet{Nao11a} presents a flipping example with a $3M_J$ planet as
the outer perturber, while the initial mutual inclination of the two
planets is up to $71.5^\circ$. \citet{LN11} shows that, if the outer
perturbers are in more eccentric orbits, the relative inclination
can be reduced, but still as high as $\sim 60^o$ for retrograde
motion to be occurred. Thus, the origin of such high mutual
inclination itself merits explanations.

In this paper, we propose a mechanism to efficiently excite
planetary eccentricities and inclinations with an outer residual gas
disk. After gas giants have formed and swept away the inner part of
gas disk, the residual gas disk outside will perturb the
architecture of inner planet systems. Due to the gravity of the
residual disk, vertical secular resonances would occur between the
very massive outer planet and the inner ones at some certain
locations. Then the mutual inclination between the planetary
orbits would be pumped. At this time, if the outer planet has a
non-zero inclination relative to disk midplane, which might result
from the previous planetary scattering, the mutual inclination is
possible to raise up to the Kozai critical value, then the Kozai
effect between planets would be induced. As a result, the
eccentricities and inclinations of the inner planets would be
excited to very high values.

The effect of gas disk in exciting planetary eccentricities was also
studied by \citet{Nag03, Terquem 2010, TTP12}, etc. Our present work
focus on how, with the aid of a disk, a two planet system will
execute secular resonances between them in order to trigger the
subsequent Kozai effect. We will also present the parameter studies
with a set of evolution equations. The paper is organized as
follows: in section 2, we introduce the model and two examples, and
the pumping region is displayed by scanning the $a_{1,0}-I_{2,0}$
plane. Then we point out the pumping mechanism is the secular
resonances coupled with the following Kozai resonance, and calculate
the location of secular resonances in the situation of small
eccentricities and inclinations by timescale comparisons in section
3. In section 4, we deduce the changing rates of some crucial
parameters in pumping process relative to any inertial plane, and
compare them with N-body simulation results. In section 5, influence
of planetary parameters are investigated. According to that, we give
the critical pumping conditions for a fixed gas disk. Section 6
displays situations in systems with more than two planets. Finally,
discussions and conclusions are presented in section 7.

\section{Model and Examples}

We consider a planet system with two giant planets (denote as $m_1$
and $m_2$ for inner and outer planet, respectively) orbiting around
a central star, with a protoplanetary disk whose inner part had been
swept out by giant planets \citep{Zhang08, ZZ10a, ZZ10b}. The gas
disk is assumed to be a two-dimensional circular annulus for
simplicity, with its mass distributed on the midplane. As both the
mass and the angular momentum of the disk are much larger than those
of the planets, we further suppose the gravity of the planets has no
influence on the disk, i.e., the disk is invariable. As the disk
exerts the gravity onto the planets, the equation of planet motion
can be written as follows
\begin{equation}\label{drdt}
\frac{d^2\textit{\textbf{r}}_i}{dt^2}
=-\frac{G(m_0+m_i)}{r_i^2}\Bigg(\frac{\textit{\textbf{r}}_i}{r_i}\Bigg)+\sum_{j\ne
i}^NGm_j\Bigg[\frac{\textit{\textbf{r}}_j-\textit{\textbf{r}}_i}
{|\textit{\textbf{r}}_j-\textit{\textbf{r}}_i|^3}-\frac{\textit{\textbf{r}}_j}{r_j^3}\Bigg]-\bigtriangledown
\Phi,
\end{equation}
where $\textit{\textbf{r}}_i$ is the position vector of the planets
relative to the star, and
\begin{equation} \label{potential}
\Phi=-G\int_{R_{\rm in}}^{R_{\rm
out}}\Sigma(r)rdr\int_0^{2\pi}\frac{d\phi}{(r^2+r_p^2-2rr_p\cos{\phi}\sin{\theta_p})^{1/2}}
\end{equation}
displays the gravitational potential from the disk \citep{Terquem
2010}. $R_{\rm in}, R_{\rm out}$ is the inner and outer border of
the disk. $(\alpha_p ,\varphi_p ,\theta_p)$ is the spherical
coordinates of a planet in the coordinate system settled by the star
and the disk midplane. $\Sigma(r)$ is the mass density of the disk,
and we use the most commonly exponential density distribution of the
disk radius $r$, $\Sigma(r)=\Sigma_0(r/R_{\rm out})^{-\alpha}$.
Total mass of the disk is settled by $M_{\rm disk}$ and the
expression for $\Sigma_0$ is shown in Appendix B.

\begin{figure}
\epsscale{.80} \plotone{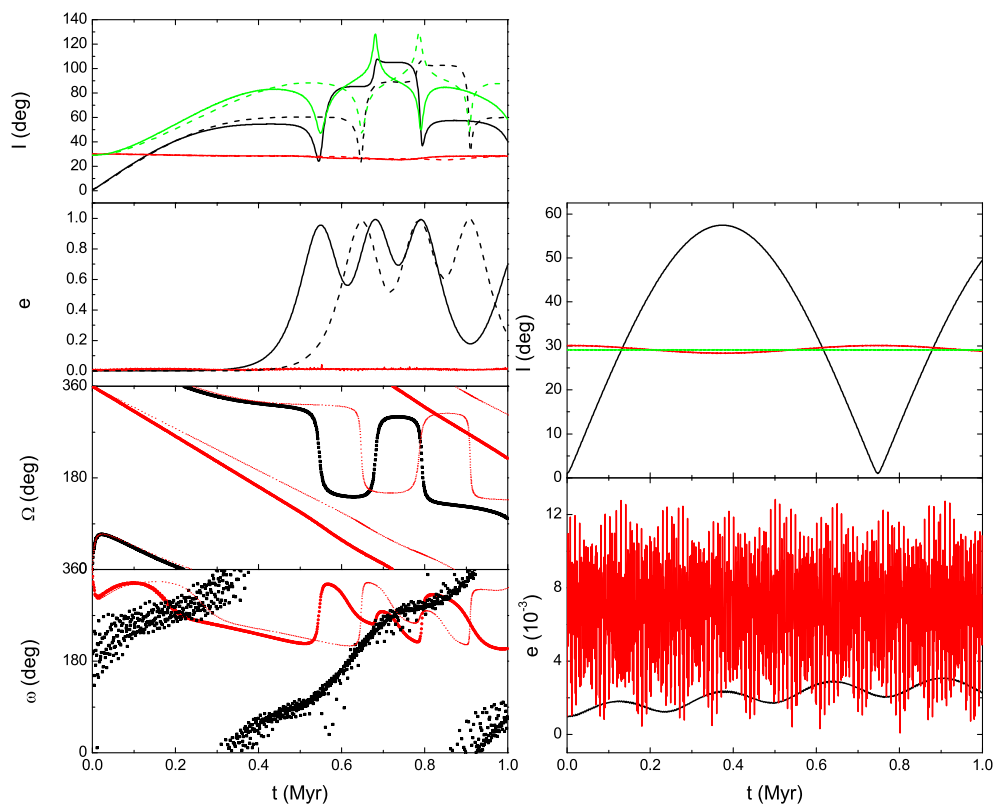} \caption{ Evolutions of two
planets with (left panels) / without (right) an outside disk's
gravity. The initial conditions are listed in Table 1. Black lines
are for the inner planet, and red for the outer one. Green lines in
the top panels indicate the mutual inclination of two planets. Dash
lines in the two left-upper panels and lighter dots in the two
left-lower panels are the results of the evolution equations
(\ref{dxdt}).  \label{special}}
\end{figure}

We apply Runge-Kutta-Fehlberg 7(8) integrator to integrate Equations
(\ref{drdt}). Figure \ref{special} gives a typical example, whose
initial conditions are listed in table 1. We set the star mass
$m_0=1M_\odot$. The inner and outer boundary of the out gas disk are
taken arbitrarily within the scope of disk observations. In order to
satisfy the assumption that the angular momentum of the disk is
overwhelming, we set the mass of the gas disk as $0.05M_\odot$.
Though it is much larger than the average mass ($0.01M_\odot$)
estimated by \citet{Williams2011}, it is still within the reasonable
range according to the recent transitional disk observation (such as
LkCa 15 \citep{Kraus2012}). The mass of outer planet is moderately
bigger than the inner one for facilitating the excitation procedure,
and the particular influence will be discussed in section
\ref{para}. We take the initial eccentricity and inclination of the
inner planet very small just to show the pumping mechanism.
Eccentricity of the outer planet is set very small in order to
conveniently compare with the results of the evolution equations
(section \ref{equa}), and the non-zero eccentricity situation will
be discussed in Section \ref{para}.

We can see from Figure \ref{special} that the inclination of the
inner planet relative to disk midplane ($I_1$) goes up to near
$50^{\circ}$ within 0.3Myr. After around 0.4Myr, the eccentricity of
the inner planet $e_1$ begins to rise, accompanied with the mutual
inclination between planets ($I_{\rm tot}$) declining.  
Ascending nodes of the two planets precess with same rates for most
of the first 0.3Myr, which implies that it is the secular resonance
that raises $I_1$ and hence $I_{\rm tot}$. This triggers the whole
excitation procedure. Argument of pericenter of the inner planet
$\omega_1$ keeps librating during the cause. We also notice that for
a while after 0.7Myr, $I_1$ becomes larger than $90^{\circ}$ and
meanwhile $e_1$ is close to 1. It provides a good opportunity for
the planet to turn into a retrograde hot-Jupiter after considering
tidal damping due to the central star. As comparison, the case with
the same initial conditions except for free of gas is shown in the
right. Eccentricities from planetary secular perturbations merely
are much smaller, and mutual inclination keeps around $30^\circ$ all
the time.

\begin{figure}
\epsscale{.80} \plotone{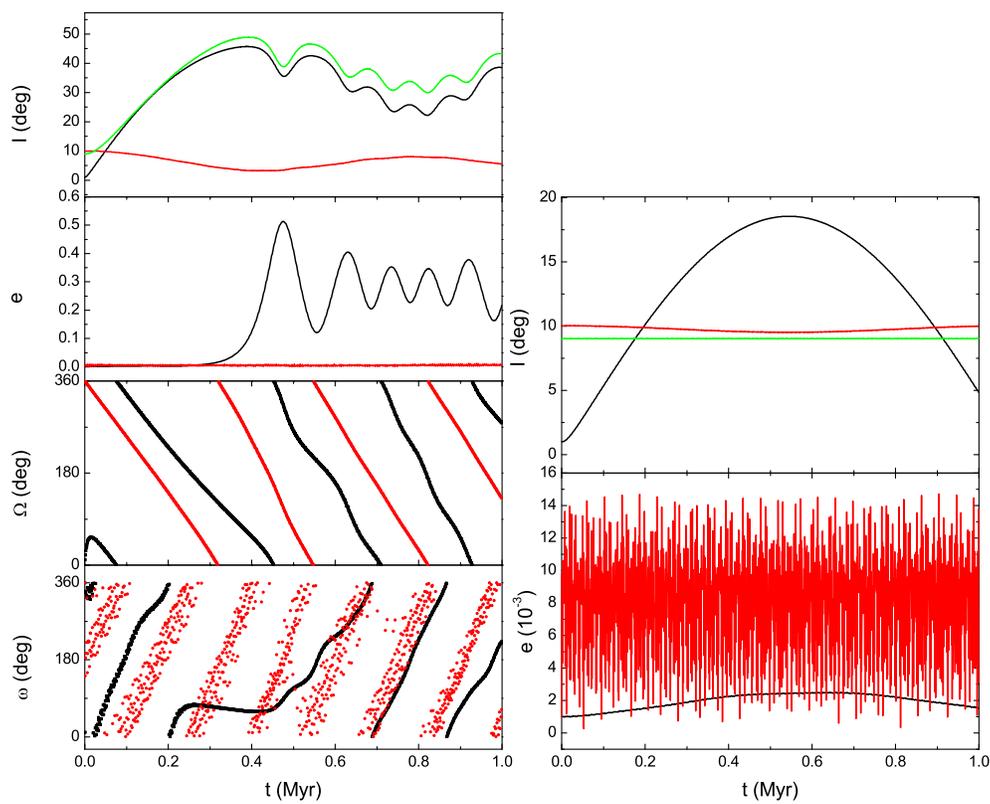} \caption{The same as Figure
\ref{special} except $a_{1,0}=5.5au, a_{2,0}=35.5au,
I_{2,0}=10^\circ$. \label{special2}}
\end{figure}

Figure \ref{special2} gives another example with smaller initial
inclination of $m_2$ ($I_{2,0}=10^\circ$)(The subscript 0 means the
initial value, and hereafter). And the mutual inclination of two
planets could also be stirred up to $40^\circ$ companied by the
approaching nodes precession rates of the two planets. Then $e_1$ is
pumped by Kozai effect with the sign of $\omega_1$ librating.

The initial inclination $I_{2,0}$ and semi-major axes are critical
parameters for pumping occurring, so we scanned the phase space of
$a_{1,0}-I_{2,0}$, and for every case, extracted the maximum of
$I_{\rm tot}$, $I_1$, $e_1$ and $e_2$ (short by $I_{\rm tot,max}$,
$I_{\rm1,max}$, $e_{\rm1,max}$ and $e_{\rm2,max}$ hereafter) during
the evolutions within 1Myr. Figure \ref{scan} in filled color shows
those with parameters the same as Figure \ref{special} except the
variable $a_{1,0}$ and $I_{2,0}$. Both $I_{\rm 1,max}$ and $I_{\rm
tot,max}$ have obvious minimums at around $a_{1,0}=3.5au$ when
$I_{2,0}=0^\circ$. And the area above the contour line of $I_{\rm
tot,max}=40^\circ$ coincides with that above the line of $e_{\rm
1,max}=0.1$  (the discrepancy in their upper left corner is
due to a longer Kozai timescale than 1Myr, so there is no enough
time for $e_1$ to rise), which implies that eccentricity pumping is
attributed into Kozai effect after inclinations have been excited.
In Figure \ref{scan}d, the eccentricity of $m_2$ is also excited in
the region that the planetary secular resonances occur (when
$I_{2,0} < 35^o$). Although $e_{\rm 2,max}$ becomes large in some
regions either due to the secular resonance or combined with the
Kozai oscillations ($I_{2,0} \ge 35^o$) from gas disk\citep{Terquem
2010, TTP12}, it still maintains less than 0.1 in most cases, which
is the basis of simplifications in the derivation of the evolution
equations in section \ref{equa}.

These pumping cases represent a possible scenario to excite
efficiently the eccentricities and inclinations of planets when
planets are far away from each other. And the pumping critical angle
is much lower than the Kozai critical angle because of the initial
inclination excitation. From the nearly equal rates of change of
nodes of two planets we have deduced it is secular resonance that
excites the inclinations. And we will further verify that in the
next section by frequency and timescale comparisons.



\section{Conditions for secular resonances (ESR and VSR) }

Secular evolution dominates dynamics of a planetary system when
planets are far away from the star and they are not close to any
low-order mean-motion resonances. In this context, once the
precession frequencies of planets are integer multiples of each
other, secular resonance would occur \citep{Lithwick 2011,Nagasawa
2000b}.At the place where the timescales of perihelion (nodal)
processing rate of two planets are equal due to the disk and mutual
planetary perturbations, secular resonance would occur, which are
called as eccentric (vertical, respectively) secular resonance, and
shorted as ESR (VSR, receptively).

In order to obtain the timescales more explicitly, we first assume
the initial eccentricities and inclinations of both planets are
small before they are effectively excited. We further assume that
$m_2\ge m_1$, and $a_2\gg a_1$, so the evolution of $m_2$ is
dominated by perturbations from the disk, and that of $m_1$ is
mainly affected by perturbations from $m_2$ (also see Figure
\ref{5dxdt}).

Under these assumptions, we use Lagrange equations \citep{MD1999} to
derive the apsidal and nodal precession rate exerted by the disk
gravity (see Appendix B for details)
\begin{equation}\label{dOM}
\dot{\Omega}_{i,\rm disk}=\frac{3}{2n_i}K\cos{I_i},
\end{equation}
\begin{equation}\label{dso}
\dot{\omega}_{i,\rm disk}=-\frac{2}{n_i}K,
\end{equation}
where $I_i$ and $\Omega_i$ are the inclinations and ascending nodes
of the two planets ($i=1,2$) with respect to the disk midplane,
$\omega_i$ is the argument of perihelion, $n_i$ is the angular
velocity of planetary mean motion, and
\begin{equation}
K=\frac{-\alpha+2}{1-\eta^{-\alpha+2}}\frac{-1+\eta^{-1-\alpha}}{-1-\alpha}\frac{GM_{\rm
disk}}{2R_{\rm out}^3}
\end{equation}
is merely related to the disk parameters, $\eta=R_{\rm in}/R_{\rm
out}$, $\alpha$ is the exponential index of disk profile.

In deriving Equations (\ref{dOM})-(\ref{dso}), terms with $e^2$ and
$\sin^2{I}$ have been eliminated to simplify the expressions, which
is suitable before the exciting of $e$ and $I$. Meanwhile, under
these assumptions, the semi-major axis $a$, eccentricity $e$ and
inclination $I$ of each planet have no secular trend from disk
gravity (see Appendix B). So the timescales of planet apsidal and
nodal precession from disk gravity can be estimated by
$2\pi/\dot{\omega}$ and $2\pi/\dot{\Omega}$ separately. Then the
timescales of the outer planet are
\begin{equation} \label{t}
\tau_{\Omega_2}=\frac{2\pi}{\dot{\Omega}_{2,\rm disk}},\qquad\qquad
\tau_{\omega_2}=\frac{2\pi}{\dot{\omega}_{2,\rm disk}}.
\end{equation}
Moreover, we apply the secular perturbation theory \citep{MD1999} to
obtain the precession timescale of the inner planet due to planetary
interactions. There are two eigenfrequencies $g_1$, $g_2$ (where
$g_1>g_2$) for $e-\omega$ solution and one eigenfrequency $f$ for
$I-\Omega$ solution in two-planet systems. So
\begin{equation}
\tau_{\Omega_1}=\frac{2\pi}{f},\qquad\qquad
\tau_{\omega_1}=\frac{2\pi}{g_1}
\end{equation}
can be used to display the precession of $\Omega_1$ and $\omega_1$
respectively.

\begin{figure}
\epsscale{1} \plotone{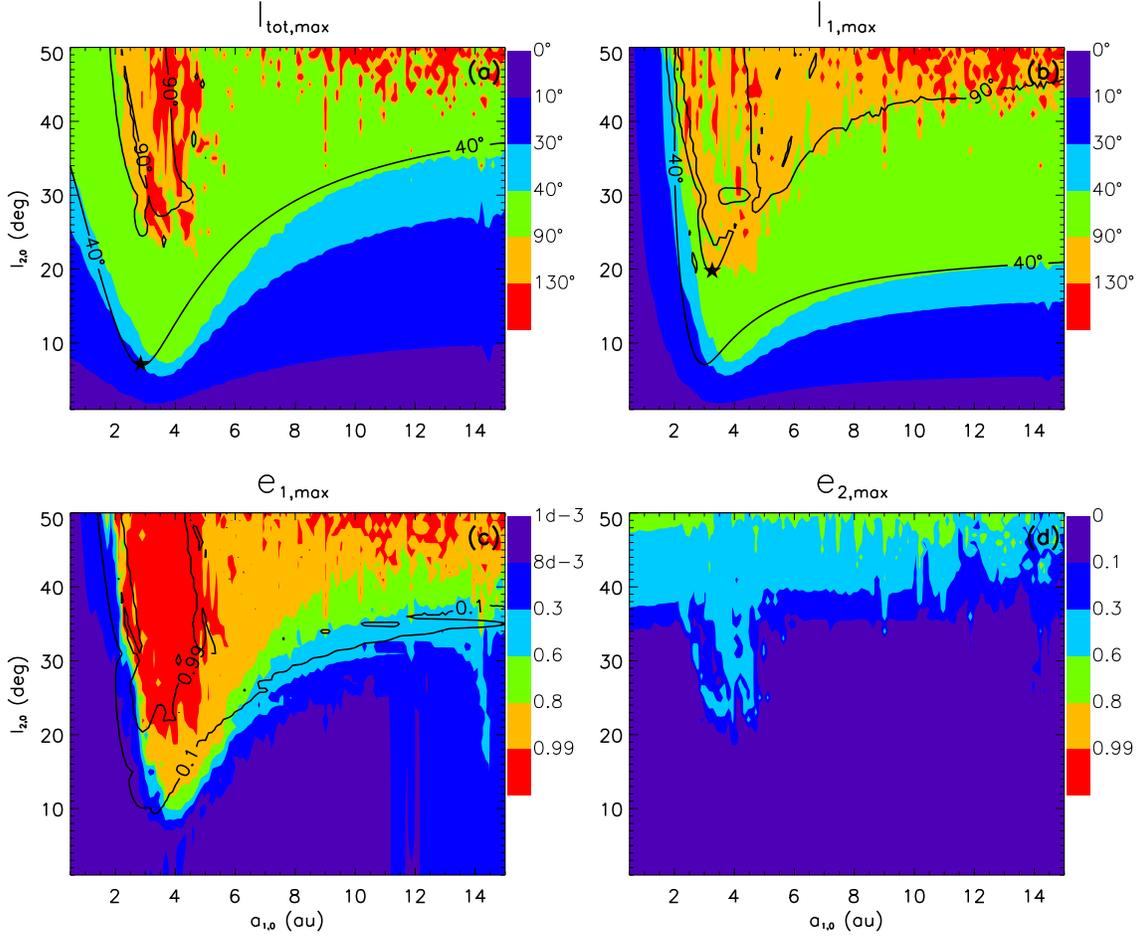} \caption{Contours of maximum of the
mutual inclination between two planets $I_{\rm tot,max}$(a), maximum
of the inclination of the inner planet $I_{\rm 1,max}$(b), maximum
of the eccentricity of the inner planet $e_{\rm 1,max}$(c), maximum
of the eccentricity of the outer planet $e_{\rm 2,max}$(d)  from
full N-body simulations during the evolution of 1Myr. Every point
has different initial inclination of the outer planet $I_{2,0}$ ($y$
axis) and different initial semi-major axis of the inner planet
$a_{1,0}$ ($x$ axis). The black lines in panel a,b and c indicate
the results of the evolution equations (\ref{dxdt}), which are
integrated 1 million years or truncated after $e_1>0.99$. The black
stars in the two upper panels are used to label the positions whose
coordinates are contoured in Figure \ref{scanscan}. \label{scan}}
\end{figure}

\begin{figure}
\epsscale{.80} \plotone{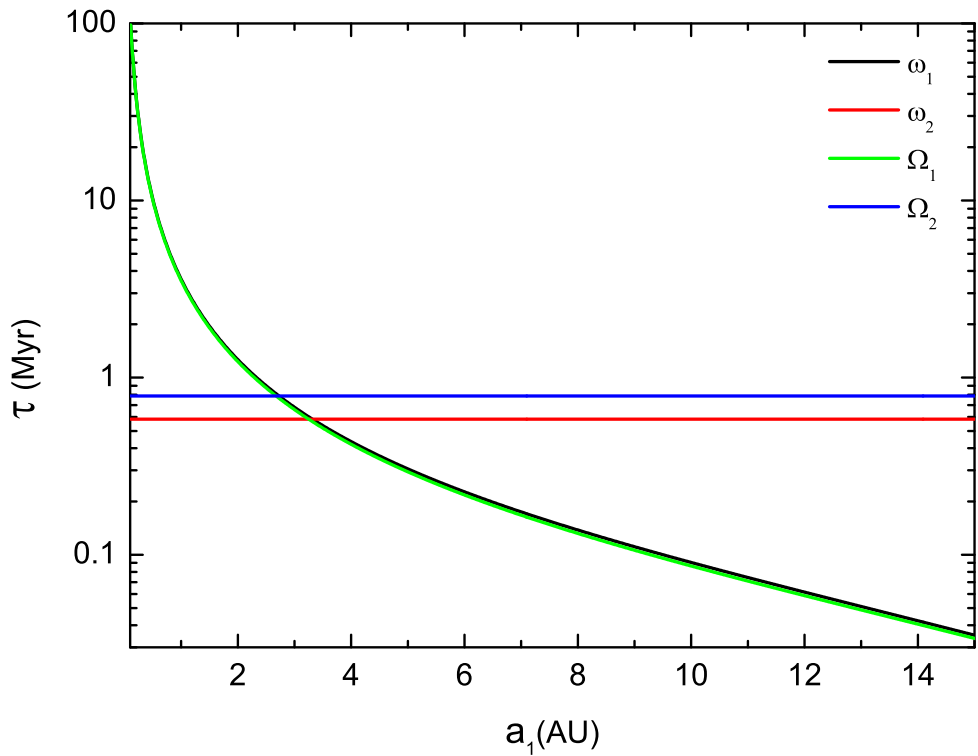} \caption{Precession
timescales of argument of pericenter $\omega$ and longitude of
ascending node $\Omega$ of two planets. Initial parameter is listed
in Table 1, except for $I_{2,0}=0$, and $a_1$ altering from 0 to 15au. Secular
resonance for $e-\omega$ would take place around 3.3au, the place
$\tau_{\omega_1}=\tau_{\omega_2}$, and secular resonance for
$I-\Omega$ around 2.7au, the place
$\tau_{\Omega_1}=\tau_{\Omega_2}$.\label{timescale}}
\end{figure}

\begin{figure}
\epsscale{.80} \plotone{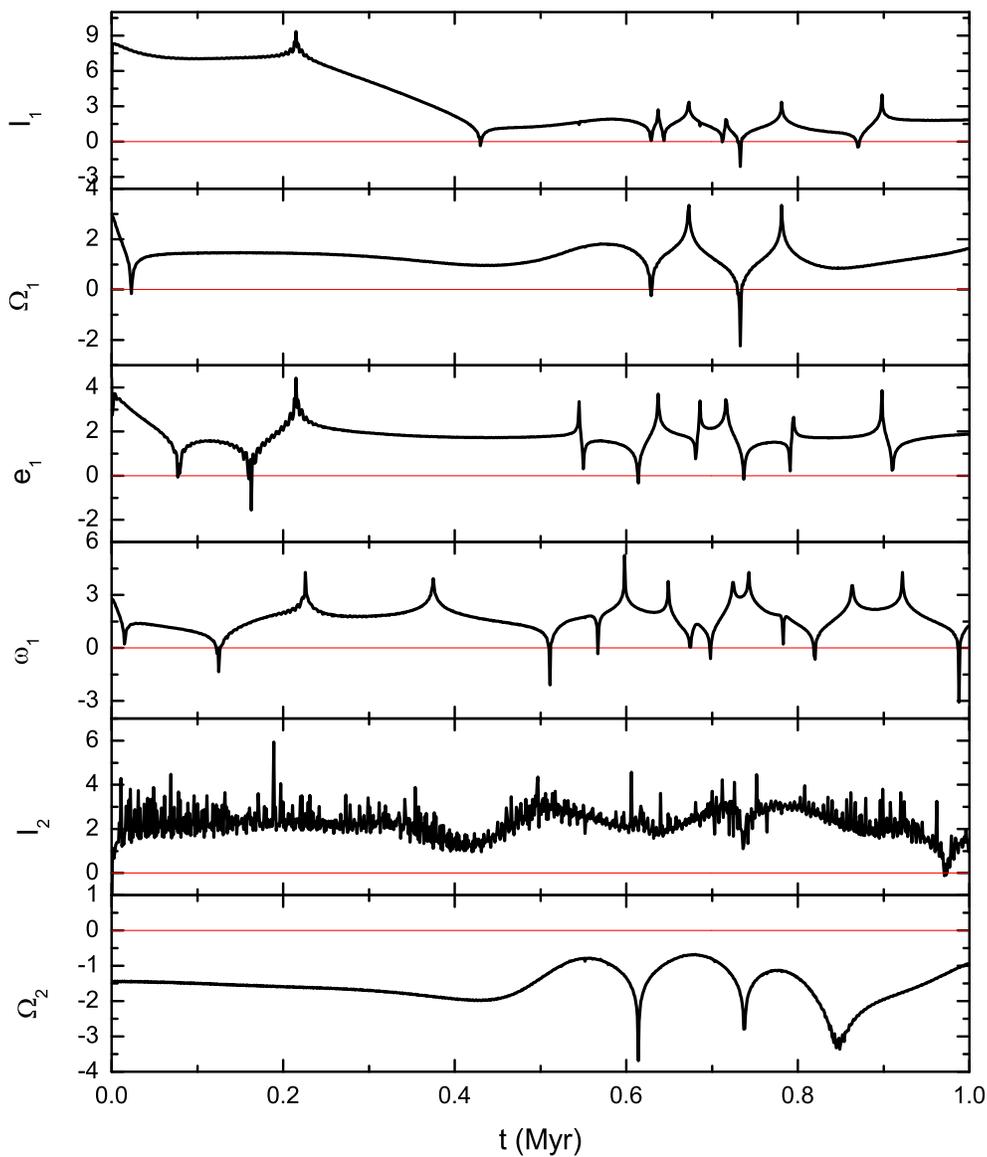} \caption{The evolution of
$\log[(dx/dt)_p/(dx/dt)_{\rm disk}]$ ($x$ represents $I_1$,
$\Omega_1$, $e_1$, $\omega_1$, $I_2$ and $\Omega_2$) with time for
the case in Figure \ref{special}. Red line is the boundary where
$(dx/dt)_p=(dx/dt)_{\rm disk}$. \label{5dxdt}}
\end{figure}

Figure \ref{timescale} shows these timescales versus the inner
planet's semi-major axis, with initial condition the same as Figure
\ref{scan} except $I_{2,0}=0$. $\tau_{\Omega_1}$ and
$\tau_{\Omega_2}$, $\tau_{\omega_1}$ and $\tau_{\omega_2}$
respectively have one cross point in Figure \ref{timescale}. The
x-coordinations of the points display the value of $a_{1,0}$ when
VSR and ESR occur, which roughly match the location of pumping at
$I_{2,0}=0$ in Figure \ref{scan}. And the y-coordinations estimate
the timescales of secular resonances, which are much less than the
average ages of gas disk \citep{Haisch2001}.

When $I_{2,0}>0$, $\tau_{\Omega_2}$ becomes larger(Equation
(\ref{dOM})), then the cross point of $\tau_{\Omega_1}$ and
$\tau_{\Omega_2}$ would move inward along the $\tau_{\Omega_1}$
line. So it only provides the estimation of the inner border of the
excitation region in Figure \ref{scan}. In order to estimate the
excitation region more precisely, we will give the evolution
equations of the elements in the next section.

\section{Evolution equations at arbitrary inclinations}
\label{equa}

To obtain the quantitative description of planetary orbits when
secular resonance happens, we will develop a set of simplified
equations to describe the evolutions of
$e_1$,$\omega_1$,$I_1,\Omega_1$, $I_2,\Omega_2$, which are suitable
for arbitrary inclinations (but still require for small $e_2$). We
set the disk midplane as the reference plane, which is assumed to
coincide with the equatorial plane of the center star. So our
derivations are different with \citet{Naoz11} in the context of
three-body systems, as their reference plane is the invariable plane
of the system.

At first, according to \citet{Mardling 2002}, the secular evolution
of the elements of $m_1$ effected by $m_2$ is expressed into the
angular momentum vector
$\textit{\textbf{h}}=\textit{\textbf{r}}\bf{\times}\dot{\textit{\textbf{r}}}$,
the Runge-Lenz vector $\textit{\textbf{e}}$ and
$\hat{\textit{\textbf{q}}}=\hat{\textit{\textbf{h}}}\bf{\times}\hat{\textit{\textbf{e}}}$
(see Equation (\ref{adedt})-(\ref{adodt}))(The hat indicates the
unit vector). And for $m_2$, the correspond vectors are
$\textit{\textbf{H}}, \textit{\textbf{E}}$ and
$\textit{\textbf{Q}}$. Then time-averaging is executed, first over
the inner orbit for eliminating eccentric anomaly $E_1$ then over
the outer orbit for removing $E_2$ (the results see Equation
(\ref{2didt})-(\ref{2dodt})). Then after, we expand the two groups
of unit vectors
$(\hat{\textit{\textbf{e}}},\hat{\textit{\textbf{q}}},\hat{\textit{\textbf{h}}})$
and
$(\hat{\textit{\textbf{E}}},\hat{\textit{\textbf{Q}}},\hat{\textit{\textbf{H}}})$
into terms with $I_1$, $\omega_1$, $\Omega_1$ and $I_2$, $\omega_2$,
$\Omega_2$ separately(see Equation (\ref{eqh})). This is the key
step to make the final formulas relative to an arbitrary plane
rather than the invariable plane of two orbits. Finally, we obtain
the evolutions of the elements due to planetary perturbation up to
the quadrupole terms, without any reductions on the eccentricities
and inclinations (see Equation (\ref{3didt})-(\ref{3dodt})). It is
worth mentioning that, the evolutions of $e_1$ and $\omega_1$ has no
assumption of $\Delta\Omega=\pi$, so has more terms than the
quadrupole parts in formula (C9) and (C5) of \citet{Naoz11}.

The disturbing from gas disk is been considered independently.
Details are in Appendix B. The final evolutions can be acquired by
adding the two parts together,
\begin{equation} \label{dxdt}
\frac{dx}{dt}=(\frac{dx}{dt})_p+(\frac{dx}{dt})_{\rm disk}.
\end{equation}
$x$ represents the six elements
$I_1$,$I_2$,$\Omega_1$,$\Omega_2$,$e_1$ and $\omega_1$. We set
$e_2=0$ as $e_2$ keeps small in most cases (Fig. \ref{scan}d), then
the six equations presented by the above one become closed (we call
them ``the evolution equations" hereafter).

We made comparisons for the two parts of the evolution equations by
drawing $\log[(dx/dt)_p/(dx/dt)_{\rm disk}]$ from true N-body
simulation in Figure \ref{5dxdt}. As was expected, for the elements
of $m_1$, $(dx/dt)_p\gg(dx/dt)_{\rm disk}$ in most time, and for
$\Omega_2$, $(d\Omega_2/dt)_p\ll(d\Omega_2/dt)_{\rm disk}$ all the
time. As for $I_2$, the influence from disk is much smaller because
of the small $e_2$ (see Equation (\ref{didtdisk})). These can be
utilized in the further deductions and simplifications.

Via the evolution equations, we can calculate the evolution of
inclination and eccentricity of the inner planet more quickly. The
dashed line in Figure \ref{special} displays the integration results
from the evolution equations. Except for some delay, both the trend
and the amplitude are fitted pretty well. And further, we use the
evolution equations to make scanning over $a_{1,0}-I_{2,0}$ (the
black contour lines in Figure \ref{scan}a-c, which is made up of the
maximums of the evolutions of 1Myr or before $e_1>0.99$ for every
case) to compare with the full N-body results. The simplified
results agree well qualitatively with the full N-body ones, except
some malposition, which mostly results from the quadrupole
approximation for disk gravity.

\section{Parameter analysis} \label{para}

\begin{figure}
\epsscale{1.1} \plottwo{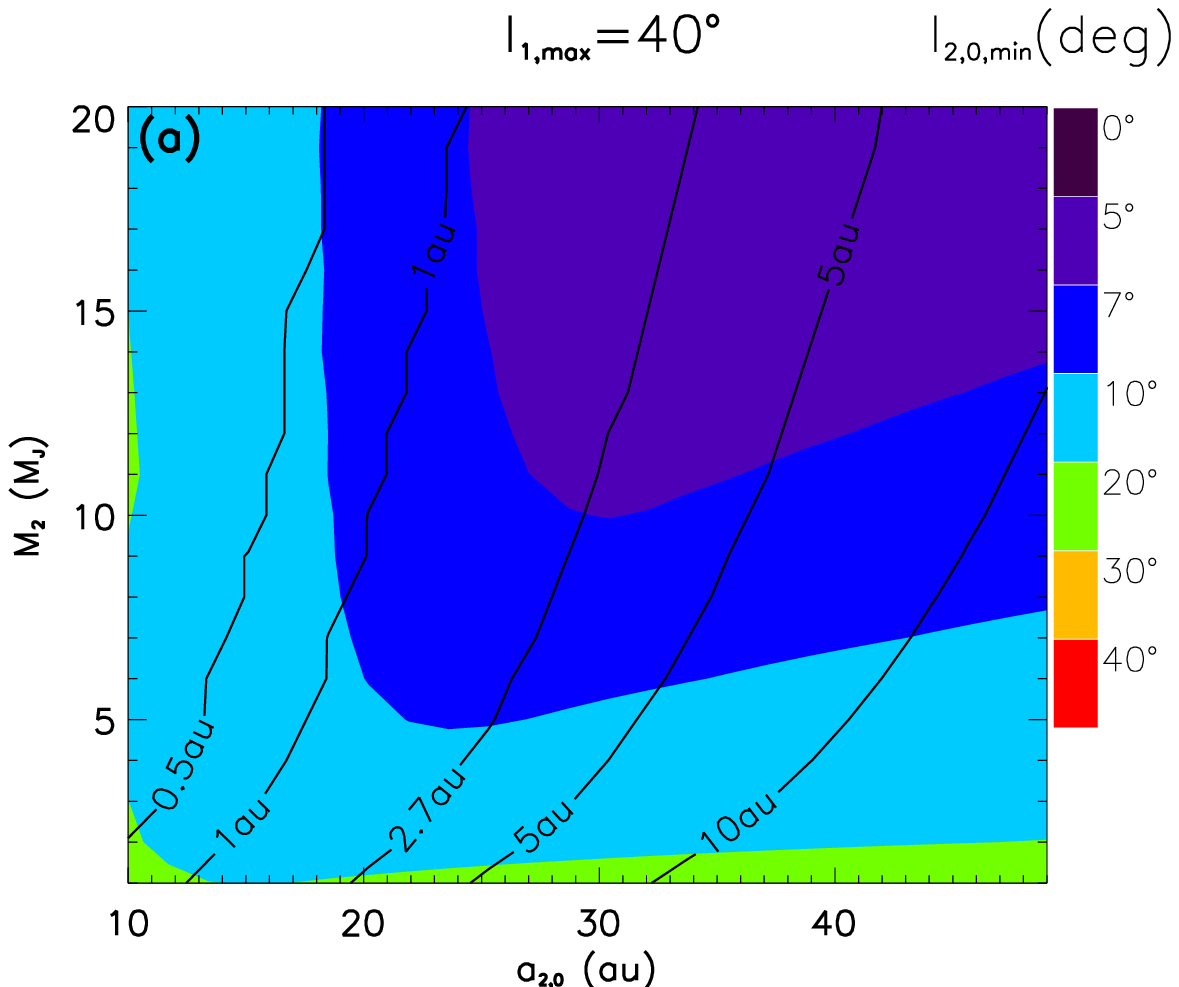}{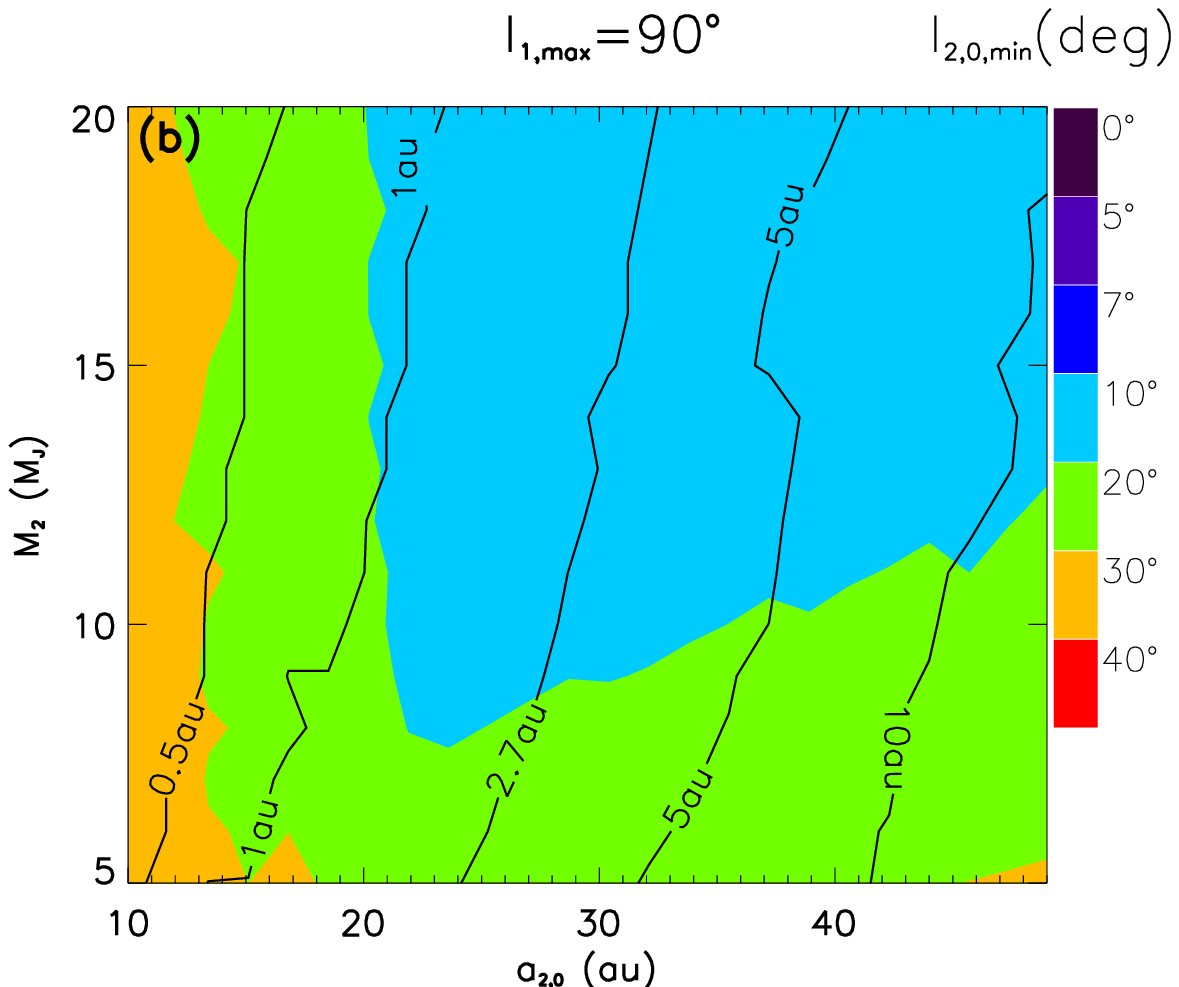} \caption{With
different $a_{2,0}$ and $m_2$, the left panel displays the minimum
of initial inclination of the outer planet (filled color contour)
for cases in which $I_{\rm tot}$ could reach $40^{\circ}$ during
evolution ($y$ coordinations of the star in Figure \ref{scan}a). The
solid line contour is made up of locations of the inner planet when
$I_{\rm tot,max}=40^{\circ}$ happens with the smallest $I_{2,0}$
($x$ coordinations of the star in Figure \ref{scan}a). The right
panel has similar meanings except  for $I_1$  reaching $90^{\circ}$
during evolution (the coordination of the star in Figure
\ref{scan}b). Every $a_{1,0}-I_{2,0}$ scanning involved is from the
same condition as the black line contours in Figure \ref{scan}.
\label{scanscan}}
\end{figure}

According to Figure \ref{scan}, Kozai resonance between planets
occurred above the contour line of $I_{\rm tot,max}=40^{\circ}$, and
the retrograde motion of $m_1$ happened above the line of $I_{\rm
1,max}=90^{\circ}$. As planets were thought to be coplanar at their
earliest stage, lower values of extremum of these two contour lines
would make the generation of retrograde motion easier. For this
purpose, we investigate the dependence of the minimums of $I_{\rm
1,max}=90^{\circ}$ and $I_{\rm tot,max}=40^{\circ}$ on $a_{2,0}$ and
$m_{2}$ with the evolution equation (\ref{dxdt})(Fig.
\ref{scanscan}), with the parameters the same as Figure
\ref{timescale} except for the variables $a_{2,0}$, $m_2$ and the
scanned $I_{2,0}$, $a_{1,0}$. Filled color contour is composed of
the values of y-coordinate of the extremum ($I_{2,0,\rm min}$),
which means the smallest inclination of $m_2$ for the onset of Kozai
effect ($I_{\rm tot,max}=40^{\circ}$) or for $m_1$ retrograding
($I_{\rm 1,max}=90^{\circ}$). The solid line contour is built up by
x-coordinates of the extremum, which signify the locations of $m_1$
when VSR between planets will occur.

Considering a Jupiter-mass planet most probably formed outside the
snowline (2.7au for a $1 M_\odot$ star, see \citealt{Ida 2004}), we
constrain the interesting scope beyond 2.7au for solid contour in
Figure \ref{scanscan}. We can see that with a $0.05 M_\odot$ gas
disk ranging from 50au to 1000au, a Jupiter-mass planet at $\sim
2.7$au will be pumped by Kozai effect with a $5m_J$ planet at 25au
and inclined $>10^{\circ}$ relative to the disk midplane. Further
more, it can be flipped into a retrograde orbit by a giant planet at
$\sim 25$ au with mass of $5m_J$ and inclination $>30^{\circ}$, or
mass of $10m_J$ with inclination $>20^{\circ}$. As every
$a_{1,0}-I_{2,0}$ scanning involved in Figure \ref{scanscan} is made
up of the $\le 1$ Myr integrations of the evolution equations, a
general gas disk aged several million years is enough for the
excitation process.

\begin{figure}
\epsscale{1.1} \plottwo{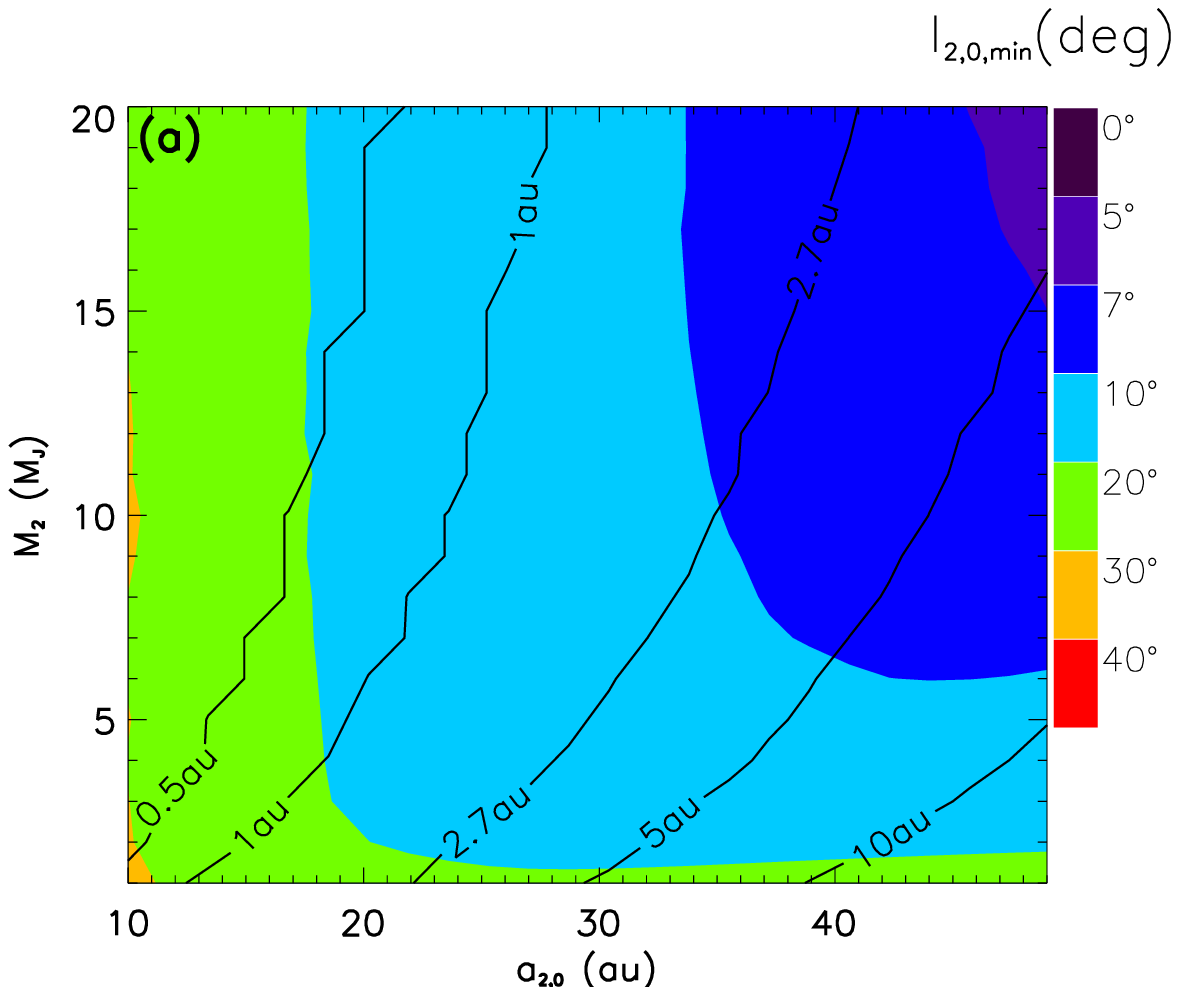}{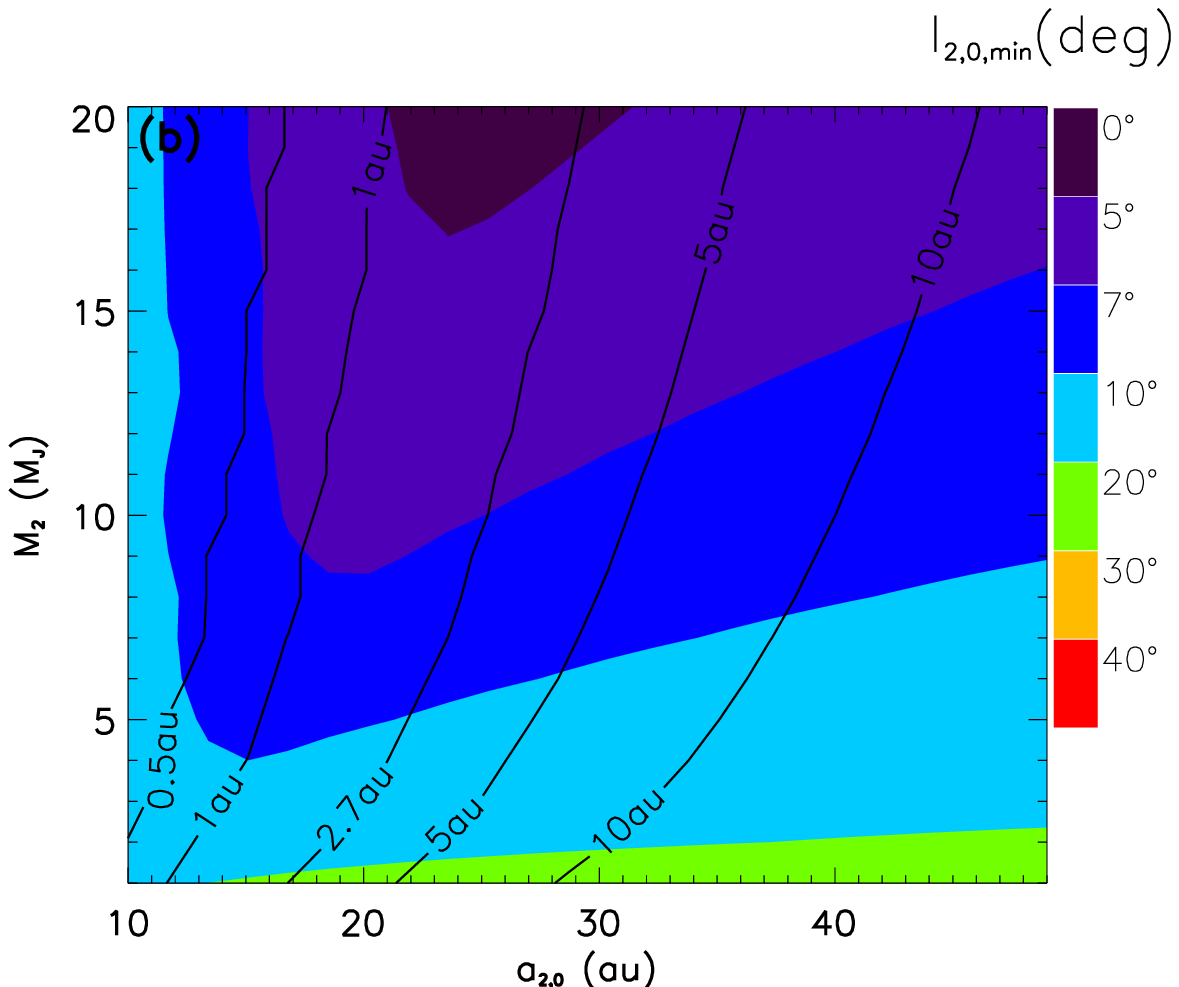} \caption{The same
as Figure \ref{scanscan}a, except for (a) $M_{disk}=20M_J$. (b)
$M_{disk}=100M_J$. \label{discmass}}
\end{figure}

We also investigate the affection of disk mass on the
exciting process. Figure \ref{discmass} gives the scanning results
like figure \ref{scanscan}a for $M_{\rm disk}=20M_J$ and $M_{\rm
disk}=100M_J$. Other parameters keep the same as in table 1 for
simplicity. For the smaller disk mass (Fig.\ref{discmass}a), the
regions of $I_{2,0,\rm min}$ move toward upper and right relative to
the same ones in figure \ref{scanscan}a, which causes the zone of
lower $I_{2,0,\rm min}$ smaller. And for the bigger disk mass
(Fig.\ref{discmass}b), the $I_{2,0,\rm min}=5^\circ \sim 7^\circ$
range extends  to the less massive $M_2$-region as compared  to
figure \ref{scanscan}a, and the $I_{2,0,\rm min}=0^\circ \sim
5^\circ$ range appears in the upper. So a more massive disk is in
favor of the pumping to some extent.

\begin{figure}
\epsscale{1.1} \plotone{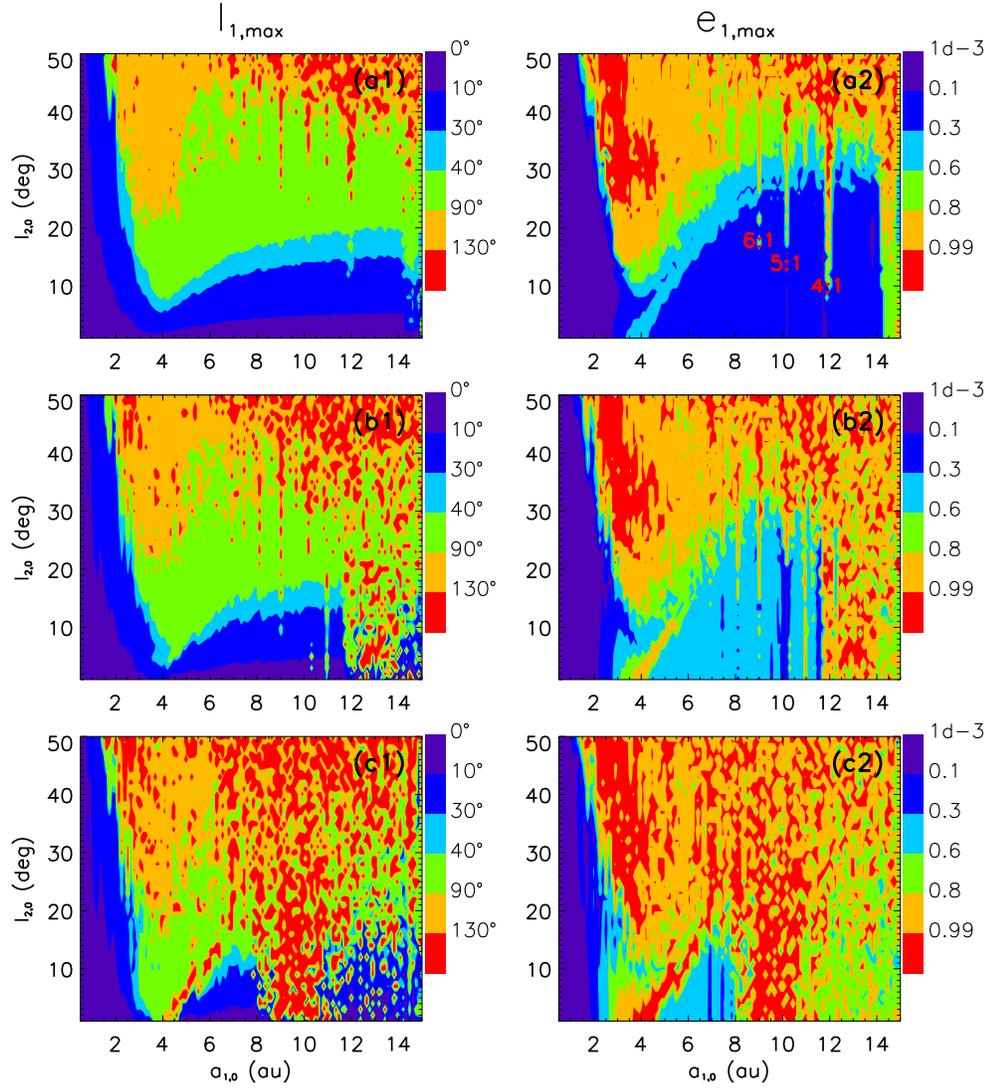} \caption{The same as the results
of N-body simulation in Figure \ref{scan}a (the left ones) and
\ref{scan}c (the right ones), except for (a) $e_{2,0}=0.2$. (b)
$e_{2,0}=0.35$. (c) $e_{2,0}=0.5$. The red numbers in the panel (a2)
mean the period ratios of two planets. \label{scane3}}
\end{figure}

All the above discussions have set $e_{2,0}=0.001$ for concentrating
on VSR more conveniently. However, $m_2$ is more likely on an
eccentric orbit since the planetary scattering have prompted a
non-zero inclination. Figure \ref{scane3} shows the same N-body
simulations scanning as those in figure \ref{scan}b,c with a higher
$e_{2,0}$. The remarkable difference in the higher $e_{2,0}$
situation is that the critical value of $I_{2,0}$ for pumping gets
smaller. The cases with $I_{1,\rm max}>90^\circ$ and $e_{1,\rm
max}>0.99$ appear even when $I_{2,0}=0^\circ$, which might be due to
the strong couplings between ESR and VSR when $e_2$ is large.
However, the effect is not obvious when $e_{2,0}<0.2$.

In Fig.8, we also notice that at small $e_{2,0}$ (Fig.8a), planetary
mean motion resonances (4:1, 5:1,6:1) cause an increase of $I_{1,\rm
max}$ and $e_{1,\rm max}$. When $e_{2,0}$ becomes larger, the
regions outside 12AU in Fig.8b and 8AU in Fig.8c are full of the
cases with $I_{1,\rm max}>130^\circ$ and $e_{1,\rm max}>0.99$, this
is due to that, as the apohelion of the inner planet is comparable
to the perihelion of the outer planet, then planetary scattering
dominates. We stop the simulation as long as any planet crosses the
inner edge of the disk.

\section{Systems with more than two planets}

\begin{figure}
\epsscale{.80} \plotone{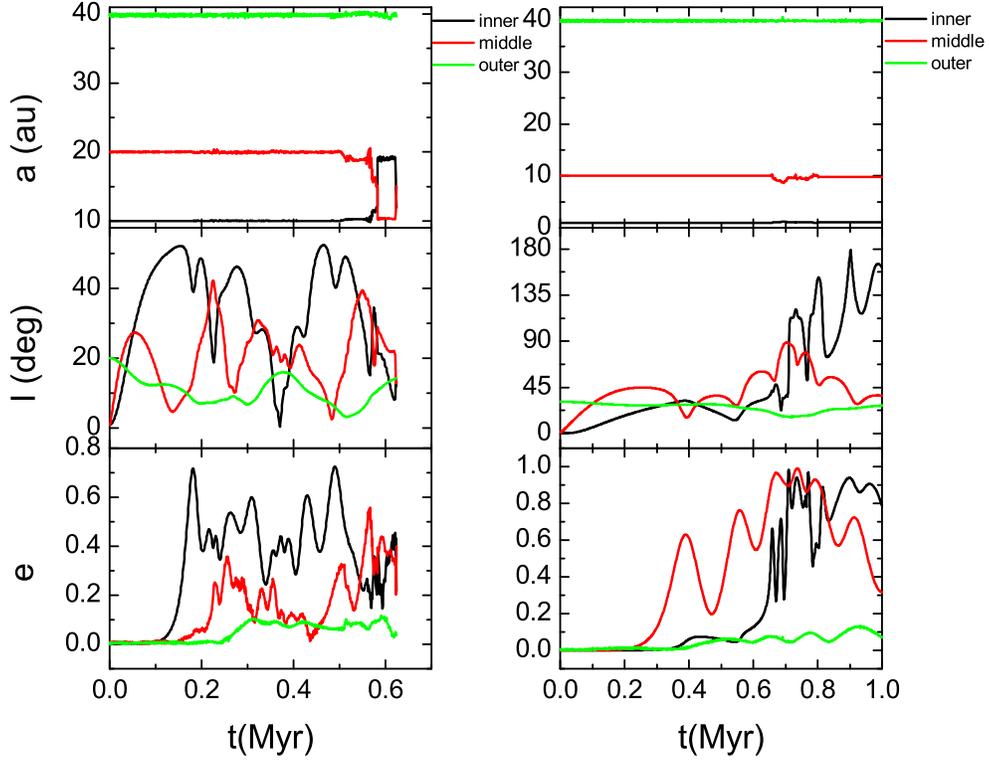} \caption{Two cases of evolution
of semi-major axis, inclinations and eccentricities of three
planets, which orbit the center star with a disk outside. The left
plot has three planets with $m_1=1m_J,m_2=1m_J,m_3=5m_J,a_1=10{\rm
au},a_2=20{\rm au},a_3=40{\rm
au},I_1=1^\circ,I_2=1^\circ,I_3=20^\circ$, and three planets in the
right plot are $m_1=0.1m_J,m_2=1m_J,m_3=5m_J,a_1=1 {\rm
au},a_2=10{\rm au},a_3=40{\rm
au},I_1=1^\circ,I_2=1^\circ,I_3=30^\circ$, The disk parameters are
the same as Table 1. \label{chain}}
\end{figure}

With the help of the VSR, a mutual inclination between planetary
orbits much smaller than the Kozai critical value can induce the
pumping of the inner planet's eccentricity eventually. However, the
occurrence of VSR constrains the inner orbit to a narrow trigger
range, and the opportunity is small that two adjacent planets happen
to be in the VSR configuration. Actually, the above pumping
mechanism can be extended to multiple planetary systems so that a
wider trigger range can be achieved. Here we give two examples to
show that the mechanism can also occur between two nonadjacent
planets, as well as inspire a chain reaction among more than two
planets. In the left case of Figure \ref{chain}, the innermost and
outermost planets were right in a configuration to be excited in a
two-planet situation, and after another planet is added between
them, the excitation still turns up. The right case in Figure
\ref{chain} exhibits a chain reaction. The middle planet is located
right in the VSR scope of the outermost planet, so its inclination
is pumped at first, which directly leads to the increase of the
mutual inclination of the inner two planets. At last, the innermost
planet is excited by the VSR with the middle planet. So the
influence of excitation of the outer planet can be spread to a more
inward scope by a chain reaction. We do not explore the specific
conditions or detailed influence of these more complicated
operations of the mechanism, and leave them to future works.

\section{Conclusions and discussions}\label{discuss}

In this paper, we proposed a mechanism to excite the eccentricities
and inclinations of planets with a residual gas disk outside the
planets. The excitation was the results of a coupling of secular
resonance and Kozai effect. After several giant planets formed, the
inner disk was assumed to have been swept out by gas giants during
their accretion, and the outer part of gas disk would coexist with
planets as long as million years. If the outermost planet has a
moderately inclined orbit relative to the disk midplane, vertical
secular resonance would happen between the planets. Then the mutual
inclination between two planets increases. Once it reaches up to
$\sim40^{\circ}$, the Kozai effect between the planets would be
induced, which can further pump the inner planet's eccentricity and
inclination to high values (Fig. \ref{special}). So this kind of
mechanism is probably one of origins of hot-Jupiters on misaligned
even retrograde orbits.

To describe the evolution of inclinations and longitude of ascending
nodes, we derived the evolution Equations, which are closed with the
assumption of $e_2=0$, and are suitable for arbitrary inclinations.
They are used to find out the locations and minimum initial
inclinations for the occurrence of vertical secular resonance and
Kozai resonance (Fig. \ref{scan}, \ref{scanscan}). The elements here
are relative to the disk midplane, and the formulas are different
with those of the elements relative to the invariable plane of the
two orbits. So they can be utilized to situations with the elements
relative to any invariable plane.

From the evolution equations, we showed that, with a mass of $0.05
M_\odot$ residual gas disk located from 50au to 1000au, a
Jupiter-mass planet will be pumped by an outer gas giant with $5m_J$
mass and $10^{\circ}$ relative to the disk midplane at least,
located out of 25au. And it could be flipped into a retrograde orbit
by an outer gas giant with $10m_J$ mass and an initial inclination
of $20^{\circ}$. Such a mechanism can be also effective for a system
with more than two planets, and the critical angles required might
be more flexible with the presence of more planets.

We used a simple disk model in order to compare with the results of
the evolution equations and fully discuss the effect of planetary
parameters. Also the limitation is that the disk mass have to be
much bigger than the total mass of planets for satisfying the
angular-momentum-advantage assumption(section 2), which restricts
the full discussion to the disc parameters. To verify that the
pumping process is irrelative to disk model, we made the same
simulations using a different disk model \citep{Nagasawa 2000a,Gang
2011}. Then we found the pumping still exists with similar
structures and even locations of contour lines in Figure \ref{scan}.

\begin{figure}
\epsscale{.80} \plotone{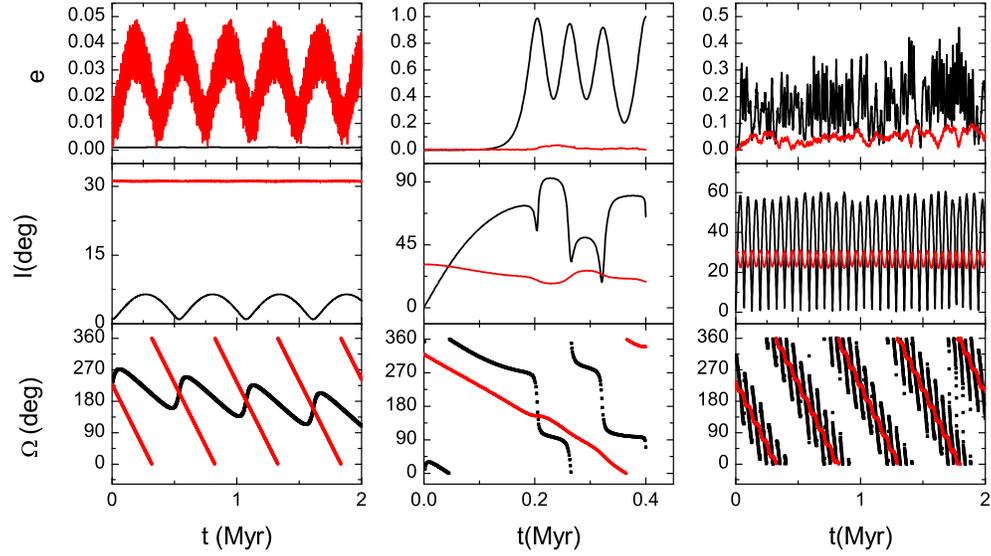} \caption{These are three cases
representing three different kinds of evolution of the inner planet.
The only different initial condition is the semi-major axis of the
inner planet $a_{1,0}$, which is 0.56{\rm au}, 4.33{\rm au} and
9.4{\rm au} from left to right. Other parameters are the same,
$m_1=1m_J, m_2=5m_J, m_{\rm disk}=50m_J, a_2=20{\rm au}, R_{\rm
in}=30{\rm au}, R_{\rm out}=1000{\rm au},
I_1=1^{\circ},I_2=31^{\circ},e_1=e_2=0.001, \Omega_1=\Omega_2$. All
arguments are arbitrary. Black lines are for the elements of the
inner planet and red lines for these of the outer planet.
\label{3case}}
\end{figure}

The mechanism revealed above has some resemblance with that in
binary systems. In a binary system with two planets orbiting one of
the stars, \citet{Takeda 2008} divided three distinct dynamical
classes according to differential nodal precessions of the two
planets. The mechanism illustrated in our paper is similar to the
so-called ``weakly coupled systems", with the same peculiarity that
the planetary mutual inclination is excited by the secular resonance
between the planets. That is actually a transitional case between
``decoupled systems" and ``dynamically rigid systems". We illustrate
this from the three cases in Figure \ref{3case}, where the semi-axis
of the inner planet is the only varying parameter. In the left case,
the  planetary secular interaction is very weak and suppressed by
the perturbation from the disk, so the secular nodal precession of
the inner planet is much slower than that of the outer planet. In
the right case, the mutual effects between the planets become so
strong that their nodal precesses coupled, and the maximum of their
mutual inclination is roughly the sum of $I_{1,0}$ and $I_{2,0}$.
The middle case is an exciting one, which occurs when planetary
interaction is big enough that the secular nodal precessions of the
two planets are approaching but not too big that the planets are
coupled.

In this respect, a protoplanetary disk has a comparable effect with
a stellar companion. Actually, this kind of analogy has been
mentioned in \citet{Wu11}. They pointed out that the place of a
planet in secular interactions could be replaced by a mass wire made
by spreading the planet along its orbit. Since protoplanetary disks
are universal in single-star systems, our exciting mechanism induced
by secular resonance would not be limited in binary systems but can
be extended to single-star systems.

Though in our mechanism, eccentricity pumping can occur with an
initial mutual inclination much smaller than the Kozai critical
angle, it is still within a rather narrow range of disk and planet
configurations for a Jupiter-mass planet can be flipped. The
efficiency for the occurrence of this mechanism in different systems
will be investigated in future works. Comparing to observations, the
narrow range also implies that, firstly, there should be most of
systems owing planets with moderate or low eccentricities and
inclinations than the systems owing retrograde hot-Jupiters.
Secondly, according to our additional simulations, the more massive
the inner planet is, the higher the initial outer inclination is
demanded to be, meanwhile, the more massive the outer planet needs
to be. So we speculate that the proportion of misalignment in
Earth-like or Neptune-like planets is probably larger than that in
Jupiter-like planet. All these need to be verified by further
statistics of simulations as well as observations.



\acknowledgments

The authors thank the referee for good suggestions which greatly
improved this paper. The work is supported by National Basic
Research Program of China (2013CB834900), Natural Science
Foundations of China (10833001, 10925313), and Fundamental Research
Funds for the Central Universities.  






\appendix
\begin{center}
{\bf APPENDIX}
\end{center}
\section{Evolution of the orbital elements due to planetary perturbation}

We apply Legendre polynomials expansion and Runge-Lenz vector
introduced in \citet{Mardling 2002} to deduce the elements'
evolution due to planetary interactions. The quadrupole contribution
of the acceleration of the inner orbit produced by the third body is
\begin{equation} \label{aquin}
\textit{\textbf{f}}_{1,p} =
\frac{Gm_2}{R^3}(3x\hat{\textit{\textbf{R}}}-\textit{\textbf{r}}),
\end{equation}
and that of the outer planet from the inner one is
\begin{equation} \label{aquout}
\textit{\textbf{f}}_{2,p}=-\frac{G\mu_{01}}{R^4}\frac{m_{012}}{m_{01}}
\Bigg[\frac{3}{2}(5x^2-r^2)\hat{\textit{\textbf{R}}}
-3x\textit{\textbf{r}}\Bigg],
\end{equation}
where $\textit{\textbf{r}}$ and $\textit{\textbf{R}}$ are position
vectors of the inner and outer planet in Jacobi coordinates,
$x=\textit{\textbf{r}}\cdot\hat{\textit{\textbf{R}}}$,
$m_{012}=m_0+m_1+m_2$, $m_{01}=m_0+m_1$, $\mu_{01}=m_0m_1/m_{01}$.

The relations between the rates of change of the inner orbital
elements and those of Runge-Lenz vectors are given by
\begin{equation}\label{adedt}
\frac{de_1}{dt}=\dot{\textit{\textbf{e}}}\cdot\hat{\textit{\textbf{e}}},
\end{equation}
\begin{equation}
\frac{d\omega_1}{dt}=-\frac{d\Omega_1}{dt}\cos{I_1}+\frac{\dot{\textit{\textbf{e}}}}
{e_1}\cdot\hat{\textit{\textbf{q}}},
\end{equation}
\begin{equation}
\frac{dI_1}{dt}=\frac{-(\sin{\omega_1}\hat{\textit{\textbf{e}}}+\cos{\omega_1}\hat{\textit{\textbf{q}}})
\cdot\dot{\textit{\textbf{h}}}}{h_1},
\end{equation}
\begin{equation}\label{adodt}
\frac{d\Omega_1}{dt}=\frac{(\cos{\omega_1}\hat{\textit{\textbf{e}}}-\sin{\omega_1}\hat{\textit{\textbf{q}}})
\cdot\dot{\textit{\textbf{h}}}}{h_1\sin{I_1}},
\end{equation}
where
\begin{equation}
\frac{d\textit{\textbf{e}}}{dt}=\frac{2(\textit{\textbf{f}}\cdot\dot{\textit{\textbf{r}}})\textit{\textbf{r}}
-(\textit{\textbf{r}}\cdot\dot{\textit{\textbf{r}}})\textit{\textbf{f}}-(\textit{\textbf{f}}\cdot\textit{\textbf{r}})
\dot{\textit{\textbf{r}}}}{Gm_{01}},
\end{equation}
\begin{equation}
\frac{d\textit{\textbf{h}}}{dt}=\textit{\textbf{r}}\times\textit{\textbf{f}}.
\end{equation}
$\textit{\textbf{h}}=\textit{\textbf{r}}\bf{\times}\dot{\textit{\textbf{r}}}$
is the orbital angular momentum vector of the inner orbit,
$\hat{\textit{\textbf{e}}}$ is the Runge-Lenz vector and
$\hat{\textit{\textbf{q}}}=\hat{\textit{\textbf{h}}}\bf{\times}\hat{\textit{\textbf{e}}}$.
$\textit{\textbf{r}}=a_1(\cos{E_1}-e_1)\hat{\textit{\textbf{e}}}
+a_1\sqrt{1-e_1^2}\sin{E_1}\hat{\textit{\textbf{q}}}$,
$\dot{\textit{\textbf{r}}}=-a_1n_1\sin{E_1}/(1-e_1\cos{E_1})\hat{\textit{\textbf{e}}}
+a_1n_1\sqrt{1-e_1^2}\cos{E_1}/(1-e_1\cos{E_1})\hat{\textit{\textbf{q}}}$.
For the outer orbit, $\textit{\textbf{r}}$ would be replaced by
$\textit{\textbf{R}}$, and the correspond unit vector is
$(\hat{\textit{\textbf{E}}},\hat{\textit{\textbf{Q}}},\hat{\textit{\textbf{H}}})$.

We separately substitute the expressions (\ref{aquin}) and
(\ref{aquout}) for $\textit{\textbf{f}}$ in
(\ref{adedt})-(\ref{adodt}), and average first over the inner orbit
then the outer orbit, and simplify the results as follow
\begin{eqnarray} \label{2didt}
(\frac{di_1}{dt})_p&=&\frac{3Gm_2}{4h_1}
\frac{a_1^2}{a_2^3}(1-e_2^2)^{-3/2}
\bigg[(\cos{\omega_1}\hat{\textit{\textbf{e}}}-\sin{\omega_1}\hat{\textit{\textbf{q}}})
+e_1^2(4\cos{\omega_1}\hat{\textit{\textbf{e}}}+\sin{\omega_1}\hat{\textit{\textbf{q}}})\bigg] \nonumber\\
&&\cdot\bigg[(\hat{\textit{\textbf{h}}}\cdot\hat{\textit{\textbf{E}}})\hat{\textit{\textbf{E}}}+
(\hat{\textit{\textbf{h}}}\cdot\hat{\textit{\textbf{Q}}})\hat{\textit{\textbf{Q}}}\bigg],
\end{eqnarray}
\begin{eqnarray}
(\frac{d\Omega_1}{dt})_p&=&\frac{3Gm_2}{4h_1\sin{i_1}}
\frac{a_1^2}{a_2^3}(1-e_2^2)^{-3/2}
\bigg[(\sin{\omega_1}\hat{\textit{\textbf{e}}}+\cos{\omega_1}\hat{\textit{\textbf{q}}})
+e_1^2(4\sin{\omega_1}\hat{\textit{\textbf{e}}}-\cos{\omega_1}\hat{\textit{\textbf{q}}})\bigg] \nonumber\\
&&\cdot\bigg[(\hat{\textit{\textbf{h}}}\cdot\hat{\textit{\textbf{E}}})\hat{\textit{\textbf{E}}}+
(\hat{\textit{\textbf{h}}}\cdot\hat{\textit{\textbf{Q}}})\hat{\textit{\textbf{Q}}}\bigg],
\end{eqnarray}
\begin{eqnarray}
(\frac{di_2}{dt})_p&=&\frac{3G\mu_{01}m_{012}}{4h_2m_{01}}\frac{a_1^2}{a_2^3}(1-e_2^2)^{-3/2}
(\cos{\omega_2}\hat{\textit{\textbf{E}}}-\sin{\omega_2}\hat{\textit{\textbf{Q}}})\cdot
\bigg[(1+4e^2_1)(\hat{\textit{\textbf{H}}}\cdot\hat{\textit{\textbf{e}}})\hat{\textit{\textbf{e}}}
\nonumber\\&&+(1-e^2_1)(\hat{\textit{\textbf{H}}}\cdot\hat{\textit{\textbf{q}}})\hat{\textit{\textbf{q}}}\bigg],
\end{eqnarray}
\begin{eqnarray}
(\frac{d\Omega_2}{dt})_p&=&\frac{3G\mu_{01}m_{012}}{4h_2m_{01}\sin{i_1}}
\frac{a_1^2}{a_2^3}(1-e_2^2)^{-3/2}
(\sin{\omega_2}\hat{\textit{\textbf{E}}}+\cos{\omega_2}\hat{\textit{\textbf{Q}}})
\cdot\bigg[(1+4e^2_1)(\hat{\textit{\textbf{H}}}\cdot\hat{\textit{\textbf{e}}})\hat{\textit{\textbf{e}}}
\nonumber\\&&+(1-e^2_1)(\hat{\textit{\textbf{H}}}\cdot\hat{\textit{\textbf{q}}})\hat{\textit{\textbf{q}}}\bigg],
\end{eqnarray}
\begin{eqnarray}
(\frac{de_1}{dt})_p&=&-\frac{15m_2a_1^3}{4m_{01}a_2^3}n_1e_1\sqrt{1-e_1^2}(1-e_2^2)^{-3/2}\bigg[(\hat{\textit{\textbf{e}}}\cdot\hat{\textit{\textbf{E}}})
(\hat{\textit{\textbf{q}}}\cdot\hat{\textit{\textbf{E}}})+(\hat{\textit{\textbf{e}}}\cdot\hat{\textit{\textbf{Q}}})
(\hat{\textit{\textbf{q}}}\cdot\hat{\textit{\textbf{Q}}})\bigg],
\end{eqnarray}
\begin{eqnarray} \label{2dodt}
(\frac{d\omega_1}{dt})_p&=&-(\frac{d\Omega_2}{dt})_{p}\cos{I_1}+\frac{3m_2a_1^3}{4m_{01}a_2^3}n_1\sqrt{1-e_1^2}(1-e_2^2)^{-3/2}\Big\{4\big[
(\hat{\textit{\textbf{e}}}\cdot\hat{\textit{\textbf{E}}})^2+(\hat{\textit{\textbf{e}}}\cdot\hat{\textit{\textbf{Q}}})^2\big]
\nonumber\\&&-\big[(\hat{\textit{\textbf{q}}}\cdot\hat{\textit{\textbf{E}}})^2+(\hat{\textit{\textbf{q}}}\cdot\hat{\textit{\textbf{Q}}})^2\big]-2\Big\}.
\end{eqnarray}
The coordinates of the Runge-Lenz vectors relative to an arbitrary
inertial plane are
\begin{equation} \label{eqh}
\hat{\textit{\textbf{e}}}= \left(\begin{array}{c}
\cos{\Omega_1}\cos{\omega_1}-\sin{\Omega_1}\sin{\omega_1}\cos{i_1}\\
\sin{\Omega_1}\cos{\omega_1}+\cos{\Omega_1}\sin{\omega_1}\cos{i_1}\\
\sin{i_1}\sin{\omega_1}
\end{array}\right),
\]
\[
\hat{\textit{\textbf{q}}}= \left(\begin{array}{c}
-\cos{\Omega_1}\sin{\omega_1}-\sin{\Omega_1}\cos{\omega_1}\cos{i_1}\\
-\sin{\Omega_1}\sin{\omega_1}+\cos{\Omega_1}\cos{\omega_1}\cos{i_1}\\
\sin{i_1}\cos{\omega_1}
\end{array}\right),
\]
\[
\hat{\textit{\textbf{h}}}= \left(\begin{array}{c}
\sin{i_1}\sin{\Omega_1}\\
-\sin{i_1}\cos{\Omega_1} \\
\cos{i_1}
\end{array}\right),
\end{equation}
and for
$\hat{\textit{\textbf{E}}},\hat{\textit{\textbf{Q}}},\hat{\textit{\textbf{H}}}$,
the formulas are similar except for switching the subscripts from 1
to 2.

Then we derived the final simplified expressions for the rates of
change of $I_1$,$I_2$,$\Omega_1$,$\Omega_2$,$e_1$ and $\omega_1$
\begin{eqnarray} \label{3didt}
(\frac{dI_1}{dt})_p&=&\frac{3m_2a_1^3n_1}{4m_{01}a_2^3}(1-e_1^2)^{-1/2}(1-e_2^2)^{-3/2}\bigg[\cos{I_1}\cos{I_2}+\sin{I_1}\sin{I_2}\cos
(\Omega_1-\Omega_2)\bigg]\nonumber\\
&&\times\Bigg\{\sin{I_2}\sin(\Omega_1-\Omega_2) +\frac{1}{2}e_1^2\bigg[(3+5\cos{2\omega_1})\sin{I_2}\sin(\Omega_1-\Omega_2) \nonumber\\
&&+5\sin{2\omega_1}(\cos{I_1}\sin{I_2}\cos(\Omega_1-\Omega_2)-\sin{I_1}\cos{I_2})\bigg]\Bigg\},
\end{eqnarray}
\begin{eqnarray}
(\frac{dI_2}{dt})_p&=&\frac{3m_0m_1a_1^2n_2}{4m_{01}^2a_2^2}(1-e_2^2)^{-2}\Bigg\{-\sin{I_1}
\sin(\Omega_1-\Omega_2)\bigg[\cos{I_1}\cos{I_2}\nonumber\\&&+\sin{I_1}\sin{I_2}\cos(\Omega_1-\Omega_2)\bigg]
+\frac{1}{2}e_1^2\bigg[-3\sin{I_1}\sin(\Omega_1-\Omega_2)
\big(\cos{I_1}\cos{I_2}\nonumber\\&&+\sin{I_1}\sin{I_2}\cos(\Omega_1-\Omega_2)\big)
+5\cos{2\omega_1}\sin(\Omega_1-\Omega_2)
\big(\sin{I_1}\cos{I_1}\cos{I_2}\nonumber\\&&-(1+\cos^2{I_1})\sin{I_2}\cos(\Omega_1-\Omega_2)\big)
+5\sin{2\omega_1}\big(\sin{I_1}\cos{I_2}\cos(\Omega_1-\Omega_2)
\nonumber\\&&-\cos{I_1}\sin{I_2}\cos{2(\Omega_1-\Omega_2)}\big)\bigg]\Bigg\},
\end{eqnarray}
\begin{eqnarray}
(\frac{d\Omega_1}{dt})_p&=&\frac{3m_2a_1^3n_1}{4m_{01}a_2^3\sin{I_1}}(1-e_1^2)^{-1/2}(1-e_2^2)^{-3/2}\Bigg\{\frac{1}{4}\sin{I_1}\cos{I_1}
\bigg[2\cos{2(\Omega_1-\Omega_2)}\sin^2{I_2}\nonumber\\&&-3\cos{2I_2}-1\bigg]
+\frac{1}{2}\cos{2I_1}\sin{2I_2}\cos(\Omega_1-\Omega_2)+\frac{1}{2}e_1^2\bigg[\cos{I_1}\cos{I_2}\nonumber\\
&&+\sin{I_1}\sin{I_2}\cos (\Omega_1-\Omega_2)\bigg]
\bigg[(-3+5\cos{2\omega_1})\big(\sin{I_1}\cos{I_2}\nonumber\\&&-\cos{I_1}\sin{I_2}\cos(\Omega_1-\Omega_2)\big)
+5\sin{I_2}\sin{2\omega_1}\sin(\Omega_1-\Omega_2)\bigg]\Bigg\},
\end{eqnarray}
\begin{eqnarray}
(\frac{d\Omega_2}{dt})_p&=&\frac{3m_0m_1a_1^2n_2}{4m_{01}^2a_2^2\sin{I_2}}(1-e_2^2)^{-2}\Bigg\{\frac{1}{4}\sin{I_2}\cos{I_2}
\bigg[2\cos{2(\Omega_1-\Omega_2)}\sin^2{I_1}-3\cos{2I_1}-1\bigg] \nonumber\\
&&+\frac{1}{2}\sin{2I_1}\cos{2I_2}\cos(\Omega_1-\Omega_2)+\frac{1}{4}e_1^2\bigg[3\sin{2I_2}\big(-\cos^2{I_1}+\sin^2{I_1}\cos^2(\Omega_1-\Omega_2)\big)
\nonumber\\&&+3\sin{2I_1}\cos{2I_2}\cos(\Omega_1-\Omega_2)-5\cos{2\omega_1}\sin{2I_1}\cos{2I_2}\cos(\Omega_1-\Omega_2)
\nonumber\\&&+5\cos{2\omega_1}\sin{2I_2}\big(\cos^2{I_1}\cos^2(\Omega_1-\Omega_2)-\sin^2(\Omega_1-\Omega_2)-\sin^2{I_1}\big)
\nonumber\\&&+10\sin{2\omega_1}\sin(\Omega_1-\Omega_2)\big(\sin{I_1}\cos{2I_2}-\cos{I_1}\sin{2I_2}\cos(\Omega_1-\Omega_2)\big)\bigg]\Bigg\},
\end{eqnarray}
\begin{eqnarray}
(\frac{de_1}{dt})_p&=&\frac{15m_2a_1^3n_1}{8m_{01}a_2^3}e_1\sqrt{1-e_1^2}(1-e_2^2)^{-3/2}\Bigg\{\sin{2\omega_1}\bigg[\big(\sin{I_1}\cos{I_2}-\cos{I_1}\sin{I_2}\cos(\Omega_1-\Omega_2)\big)^2
\nonumber\\&&-\sin^2{I_2}\sin^2(\Omega_1-\Omega_2)\bigg]-2\cos{2\omega_1}\sin{I_2}\sin(\Omega_1-\Omega_2)\bigg[\sin{I_1}\cos{I_2}
\nonumber\\&&-\cos{I_1}\sin{I_2}\cos(\Omega_1-\Omega_2)\bigg]\Bigg\},
\end{eqnarray}
\begin{eqnarray} \label{3dodt}
(\frac{d\omega_1}{dt})_p&=&-(\frac{d\Omega_1}{dt})_p\cos{I_1}-\frac{3m_2a_1^3n_1}{4m_{01}a_2^3}\sqrt{1-e_1^2}(1-e_2^2)^{-3/2}\Bigg\{\big(1-5\sin^2{\omega_1}\big)\bigg[\big(\sin{I_1}\sin{I_2}
\nonumber\\&&+\cos{I_1}\cos{I_2}\cos(\Omega_1-\Omega_2)\big)^2+\sin^2(\Omega_1-\Omega_2)\big(\cos^2{I_1}+\sin^2{I_2}\big)-1\bigg]
\nonumber\\&&-5\sin{2\omega_1}\sin(\Omega_1-\Omega_2)\sin{I_2}\bigg[\sin{I_1}\cos{I_2}-\cos{I_1}\sin{I_2}\cos(\Omega_1-\Omega_2)\bigg]
\nonumber\\&&+3\sin^2{I_2}\sin^2(\Omega_1-\Omega_2)-1\Bigg\}.
\end{eqnarray}
When $\Omega_1-\Omega_2=\pi$, the latter two formula turn to the
quadrupole parts of (C9) and (C5) of \citet{Naoz11}.

\section{Evolution of the orbital elements due to disk gravity}

As in observation, disk mass is a commonly estimated parameter
rather than the radial distribution exponential or mass density, we
set disk mass as an independent parament and deduce the mass density
from
\begin{equation}
\int_{R_{\rm in}}^{R_{\rm out}}\Sigma_0\Bigg(\frac{r}{R_{\rm
out}}\Bigg)^{-\alpha} 2\pi rdr=M_{\rm disk},
\end{equation}
then obtain
\begin{equation}
\Sigma_0=\frac{(-\alpha+2)M_{\rm disk}}{2(1-\eta^{-\alpha+2})\pi
R_{\rm out}^2}
\end{equation}
with $\eta=R_{\rm in}/R_{\rm out}$.

We used Lagrange's equations in \citet{MD1999} to deduce the rates
of change of elements due to disk gravity. First, we expanded the
gravity potential in $r_p/r$ to the quadrupole, like \citet{Terquem
2010},
\begin{equation}
\Phi=-\frac{-\alpha+2}{1-\eta^{-\alpha+2}}\frac{GM_{\rm
disk}}{R_{\rm out}}
\Bigg[\frac{1-\eta^{1-\alpha}}{1-\alpha}+\frac{-1+\eta^{-1-\alpha}}{1+\alpha}
\frac{r_p^2}{2R_{\rm
out}^2}\bigg(-1+\frac{3}{2}\sin^2{\theta_p}\bigg)\Bigg].
\end{equation}
The first term in the square brackets has no contribution to
derivation, so only the second one is retained. Defining
\begin{equation}
K=\frac{-\alpha+2}{1-\eta^{-\alpha+2}}\frac{-1+\eta^{-1-\alpha}}{-1-\alpha}\frac{GM_{\rm
disk}}{2R_{\rm out}^3},
\end{equation}
then substituted the expresses with true anomaly $f$ for $r_p$ and
$\theta_p$, we got
\begin{equation}
\Phi=K\frac{a^2(1-e^2)^2}{(1+e\cos{f})^2}\bigg[\frac{1}{2}-\frac{3}{2}\sin^2{(\omega+f)}\sin^2{I}\bigg].
\end{equation}

We substituted the above one into Lagrange's Equations
(6.148)-(6.150) in \citet{MD1999}, then averaged over true anomaly
$f$, and got the evolutions finally
\begin{equation}
(\frac{da}{dt})_{\rm disk}=0,
\end{equation}
\begin{equation}
(\frac{de}{dt})_{\rm
disk}=-\frac{15Ke\beta}{4n}\sin{2\omega}\sin^2{I},
\end{equation}
\begin{equation} \label{didtdisk}
(\frac{dI}{dt})_{\rm
disk}=\frac{15Ke^2}{8n\beta}\sin{2\omega}\sin{2I},
\end{equation}
\begin{equation}
(\frac{d\Omega}{dt})_{\rm
disk}=\frac{3K\cos{I}}{4n\beta}(2+3e^2-5e^2\cos{2\omega}),
\end{equation}
\begin{equation}
(\frac{d\omega}{dt})_{\rm
disk}=\frac{K}{n\beta}\Bigg\{-2-\frac{9}{8}e^2
+\frac{15}{4}e^2\cos{2\omega}+\sin^2{I}\bigg[\frac{9}{4}+\frac{9}{16}e^2
-\frac{15}{16}(2+e^2)\cos{2\omega}\bigg]\Bigg\},
\end{equation}
where $\beta=\sqrt{1-e^2}$.

When $i\simeq0$ and $e=0$, the expressions can be simplified into
\begin{equation}
(\frac{da}{dt})_{\rm disk}=(\frac{de}{dt})_{\rm
disk}=(\frac{dI}{dt})_{\rm disk}=0
\end{equation}
\begin{equation} \label{ts}
(\frac{d\Omega}{dt})_{\rm disk}=\frac{3K\cos{I}}{2n}
\end{equation}
\begin{equation}
(\frac{d\omega}{dt})_{\rm disk}=-\frac{2K}{n}
\end{equation}




\clearpage



\begin{table}
\begin{center}
\caption{Initial condition for Figure \ref{special}.}
\begin{tabular}{ccccc}
\tableline\tableline
Planet & Mass & Semimajor Axis & Eccentricity & Inclination\\
    & ($M_J$) & (au) &  & ($^{\circ}$)\\
    \tableline
$m_1$ & 1 & 3 & 0.001 & 1 \\
$m_2$ & 10 & 30 & 0.001 & 30 \\
\tableline\tableline
disk & Mass & $R_{\rm in}$ & $R_{\rm out}$ & $\alpha$ \\
    & ($M_J$) & ({\rm au}) & ({\rm au}) &  \\ \tableline
 & 50 & 50 & 1000 & 1 \\
\tableline\tableline
\end{tabular}
\tablecomments{Other arguments are taken arbitrarily except for
longitude of nodes of two planets equal.}
\end{center}
\end{table}

\clearpage

\clearpage

\clearpage

\clearpage






\end{document}